\documentstyle[preprint,aps,floats]{revtex}
\tightenlines
\input psfig.tex
\begin{document}
\def\ba{\begin{eqnarray}}
\def\ea{\end{eqnarray}}
\def\be{\begin{equation}}
\def\ee{\end{equation}}


\title{The General Primordial Cosmic Perturbation}

\author{Martin Bucher,\thanks{E-mail: M.A.Bucher@damtp.cam.ac.uk} %
Kavilan Moodley\thanks{E-mail: K.Moodley@damtp.cam.ac.uk} %
and Neil Turok\thanks{E-mail: N.G.Turok@damtp.cam.ac.uk}\\
DAMTP, Centre for Mathematical Sciences, University of Cambridge\\
Wilberforce Road, Cambridge CB3 0WA, United Kingdom}

\date{18 April 1999 (Revised --- 29 February 2000)}
\maketitle

\begin{abstract}
We consider the most general primordial cosmological perturbation 
in a universe filled with photons, baryons, neutrinos, and a hypothetical
cold dark matter (CDM) component within the framework of linearized 
perturbation theory. We present a careful discussion of the different 
allowed modes, distinguishing modes which are regular at early times, 
singular at early times, or pure gauge.
As well as the familiar
growing and decaying adiabatic
modes and the baryonic and CDM isocurvature modes, we 
identify two {\it neutrino isocurvature} modes. In the first,
the ratio of neutrinos to photons varies spatially but the net
density perturbation vanishes. In the second the photon-baryon plasma 
and the neutrino fluid  have a spatially varying relative 
bulk velocity, balanced so that the  net momentum density vanishes. 
Possible mechanisms which could generate the two
neutrino isocurvature modes are discussed. If one allows
the most general regular primordial perturbation, 
all quadratic correlators of observables such as the microwave background
anisotropy and matter perturbations are completely determined by
a $5\times 5,$ real, symmetric 
matrix-valued function of co-moving wavenumber. 
In a companion
paper we examine prospects for detecting or
constraining the amplitudes of the most general allowed
regular perturbations using
present and future CMB data.
\end{abstract}

\def\ob{{\Omega _{b,0}}}
\def\oc{{\Omega _{c,0}}}

\section{Introduction}
A key challenge of modern cosmology
is understanding the nature of the primordial
fluctuations that eventually led to the formation of 
large scale structure in our universe. One possibility is
that the fluctuations were
generated in a period of inflation
prior to the radiation dominated era of the hot big bang.
As inflation ended the fluctuations would then have been imprinted as
initial conditions for the radiation era on scales far beyond
the Hubble radius. The second possibility
is that the structure was generated through some 
causal mechanism operating within the standard big bang radiation and
matter eras.
In this paper we focus on the first option, that the fluctuations 
were imprinted early in the radiation era as
linear fluctuations in the metric and in the matter and radiation content.

For several good reasons the possibility that the  primordial 
perturbations were adiabatic has been the focus of most interest to date.
If the relative abundances of different particle species
were determined directly from the Lagrangian 
describing local physics, one would expect those abundance ratios to
be spatially constant because all regions of the universe would
share an identical early history, independent of
the long wavelength perturbations. The 
stress-energy present in the universe would then be characterized on large 
scales by 
a single, spatially uniform equation of state.  Such
fluctuations are termed adiabatic and are the simplest possibility
for perturbing the matter content and the geometry of the universe. 
They are also naturally predicted by the simplest inflationary 
models \cite{infpert}, although there also exist
more complex models giving  other types of 
perturbations \cite{nongauss}. 
For a recent discussion see ref.~\cite{peenew}.

Nevertheless, there is no {\it a priori} reason why the situation
could not be more complicated with abundance ratios varying from 
place to place. Perturbations of this sort are known as 
{\it isocurvature,} or sometimes {\it entropy,} 
perturbations. Most studies of isocurvature perturbations have 
examined the possibility that the primordial
perturbations were entirely isocurvature with
a vanishing primordial adiabatic component and have sought
to explore whether such pure 
isocurvature models could explain the observed structure in the 
universe \cite{pjepa}--\cite{bond}. 

In this paper we adopt a more phenomenological approach,
which we believe is now warranted by the prospect of 
upcoming precision measurements of the CMB anisotropy on 
small scales. Ground based and balloon borne telescopes
and the MAP and PLANCK satellites will 
provide very detailed measurements of the
primordial fluctuations \cite{mapp,plnck}.
Most work on how
to interpret this data has focused on parameter
estimation starting from the assumption that the initial
perturbations were generated by an inflationary model
specified by a small number of undetermined
free parameters \cite{paramest}. Adiabatic perturbations, characterized
by an amplitude and spectral index, as well as tensor
fluctuations also characterized by two parameters
are usually assumed, and based on these assumptions a host
of cosmological parameters, such as $\Omega _{total},$
$\Omega _{b},$ $\Omega _{\Lambda },$ $h,$ $N_\nu ,$ 
are to be inferred from the observed
CMB multipole moments. While the prospects of such measurements
are beguiling, they rely heavily on assumptions regarding the 
form  of
the primordial perturbations. We feel that those assumptions are 
worth checking against the data using
an approach that assumes neither inflation nor any other
favorite theoretical model. 

For learning about the fundamental physics responsible
for structure formation,
determining  whether the primordial fluctuations 
were in fact Gaussian and adiabatic is at least as 
important and interesting
as measuring the values of cosmological 
parameters.
For this purpose the relevant question is not whether 
primordial isocurvature perturbations offer
a viable alternative to adiabatic perturbations, 
which has been the focus of prior work, 
but rather how and to what extent can observations constrain
the presence of isocurvature modes. Rather than 
considering a particular isocurvature model,
it seems appropriate to consider the most
general primordial perturbation possible
without adding new physics to the hot big bang era. 
In other words, we assume a universe 
filled with neutrinos, photons, leptons and baryons,
and a cold dark matter (CDM) component and then try to
place constraints on the amplitudes of all
possible perturbation modes.

As mentioned above, we shall need to make a
simplifying assumption 
to limit the parameter
space of possible perturbations 
to reasonable size. We shall assume that
the fluctuations were indeed primordial---that is,
that the perturbation modes were excited by 
physics operating at a very early epoch preceding the 
hot radiation dominated era and that 
no new dynamics influenced the perturbations 
at late times. 
Here `late' means
well before decoupling, so that whatever decaying
modes were excited earlier had a chance to
decay before influencing the observed structure in our
cosmic microwave sky. 
This assumption excludes cosmic 
defect models (i.e., strings, global textures, etc.)~%
\cite{shellard} of structure formation 
where a detailed understanding of the dynamics of
the order parameter field is required to determine 
the unequal time correlations at late times \cite{pst,durrer}.

How should the most general perturbation be characterised?
This question has historically generated some
confusion in the literature. 
The standard terminology  refers to `growing,' `decaying,' and 
`gauge' modes, the latter referring to modes affected by
general coordinate transformations preserving the chosen gauge condition. 
In this paper as in most work on cosmological
perturbations we shall use synchronous gauge. The variables in 
other common gauges (e.g., Newtonian gauge) are merely linear
combinations of the synchronous gauge variables and their time derivatives. 
In synchronous gauge, 
the two gauge modes for scalar perturbations are easily identified. 
The first corresponds to time-independent spatial reparametrizations of the 
constant time hypersurfaces. In this mode the metric perturbation 
is constant and the matter is unperturbed. The second corresponds to
deformations of these hypersurfaces through a spatially
dependent shift in the time direction. For the latter gauge mode,
which diverges at early times, 
both the matter variables and the metric are perturbed. 
Indeed this mode may be regarded
as a shift in the time of the big bang singularity. 
The remaining modes are then defined modulo the two gauge modes. 

We characterise the remaining modes as either `regular' or `singular,'
according to their behavior as the time since the big bang tends to
zero. 
This terminology is preferable to the standard one because it includes 
constant modes such as the neutrino velocity mode we shall discuss.
In this paper we shall treat only the `regular' modes (i.e., those
regular up to the gauge modes).
There are several reasons for this. 
Singular modes are necessarily decaying as one proceeds
forward in time away from the big bang, so if they are present 
at some very 
early time with linear amplitude so that a perturbative treatment
is valid they quickly become irrelevant.  A second reason is that
even if the perturbation amplitude is small 
the perfect fluid approximation breaks down at early times
for the singular modes. Higher moments in the Boltzmann hierarchy become
progressively more important as one goes back in time,
and specifying the initial
conditions involves specifying an infinite number of constants. 
The perfect fluid approximation seems essential 
to a simple specification of primordial initial perturbations. 
Of course, decaying modes might be produced by
some late time physics (such as cosmic defects) operating well 
after the big bang, but our point here is that it would appear very difficult
to characterise them without introducing explicit source terms. 
In contrast, the `regular' modes are completely
characterised by specifying the leading terms in a power series expansion
in conformal time of the low moments of the phase space density---that is,
the density and velocity of the fluids.

Having characterised the regular perturbations by the leading terms
in the power series for the metric and fluid densities and velocities,
any quadratic observable (e.g., the matter power spectrum or the 
cosmic microwave anisotropy power spectrum) is then 
completely determined by a primordial power spectral matrix,
which 
rather than a single function of $k$ as it would be for growing mode
adiabatic perturbations is
a $5\times 5,$ real, symmetric matrix function of $k.$
The off-diagonal elements establish correlations between
the modes. As long as only quadratic observables are considered,
no assumption of Gaussianity is required. 

The counting arises as follows. Each cosmological fluid is described by
two  first order equations, so each fluid introduces two new perturbation modes.
However, in synchronous gauge, as mentioned, there is one gauge mode affecting
the fluid perturbations.
For a single fluid there is therefore just one regular,
growing mode perturbation, and no physical (i.e., non-gauge) decaying mode.
A convenient way to deal with the gauge mode is to identify it with
the velocity of the cold dark matter. By a coordinate choice this may be
chosen initially to be zero. If it is assumed that 
the cold dark matter couples to the other fluids only via 
gravity, there is no scattering term
to consider, and with coordinates chosen so that the velocity 
is initially zero, it remains so for all times.
If we now introduce photons and baryons, four new modes arise.
The first is an adiabatic decaying mode. There are also the
baryon isocurvature and cold dark matter isocurvature modes, where
the initial conditions contain equal and opposite perturbations of
the photon density and the baryon or cold dark matter densities, respectively.
The fourth new mode is that in which the photon and baryon fluids 
have a relative velocity that diverges at early times and the 
fluid approximation breaks down. So far we have three regular modes. 
Let us now introduce neutrinos into the picture. We shall imagine we
are setting up the perturbations after neutrino decoupling ($T\sim $MeV,
$t \sim $ seconds).
For our purposes there is no distinction between the different neutrino species
nor between neutrinos and antineutrinos, since we are only interested in how
perturbations in the neutrino fluid affect cosmological observations
of the density and microwave background today. Two new 
perturbation modes are introduced,  the first a neutrino isocurvature density
perturbation and the second a neutrino  isocurvature velocity perturbation.
In the latter, we can arrange the neutrino and photon-baryon 
fluids to have equal and opposite momentum density. In the approximation that
we ignore the collision term coupling 
neutrinos to  the photon-baryon fluid, which is valid
after neutrino decoupling, and in the small time limit 
there is no divergence in this mode. 

The neutrino isocurvature velocity perturbation may 
be considered as primordial as long as one takes primordial to
mean `generated after one second' but well
before photon-baryon decoupling. Of course similar remarks apply to the
cold dark matter, baryon isocurvature or neutrino isocurvature perturbations.
Namely if we go back far enough in cosmic history, where the various 
conservation laws for baryon or lepton number break down, or where
the cold dark matter was initially
generated  or reached thermal equilibrium, 
then a description of the form we are using would also 
break down. Since one second is still quite early, and 
certainly well before
photon-baryon decoupling, we regard it admissible to consider
the possibility of `primordial' neutrino velocity perturbations.
Possible mechanisms  for generating them shall be
discussed in a separate paper \cite{ltter}.

A primeval baryon isocurvature model (PBI) 
was introduced by Peebles \cite{pjepa}
in which a universe with just baryons, radiation, and neutrinos
is assumed and primordial perturbations in the baryon-to-photon
ratio are assumed. Since at early times the baryons contribute
negligibly to the total density, such perturbations lack an
adiabatic, or curvature, component at early times.
Compared to an $\Omega =1$ CDM model PBI gives:
(1) lower small-scale relative peculiar velocities,
(2) greater large-scale flow velocities,
(3) earlier reionization,
and (4) earlier galaxy and star formation \cite{cen}.
The consequences of PBI and comparison with observations 
have been studied by a number of 
authors \cite{yo,sugi,gor,chiba,hu,nnewa}.

Bond and Efstathiou \cite{bond,bondb} have considered a CDM 
isocurvature model in which the CDM-to-photon ratio
varies spatially. A possible mechanism for generating
such perturbations arises in models with axion dark matter
in which scale invariant fluctuations of the axion field
$A(x)$ are converted into density fluctuations after
the axion field acquires a mass in the QCD phase 
transition \cite{axenides}.
Under the assumption that quantum fluctuations during
inflation impart a scale free spectrum of fluctuations
to the axion field, so that $\delta A(k)\sim k^{-3/2},$ 
and that $\delta A(k)\ll \bar A,$ a scale-free spectrum 
of Gaussian fluctuations in the axion-to-photon ratio
is generated. The resulting spectrum
of density fluctuations today has the same power
law on large scales as for adiabatic fluctuations with 
a scale free spectrum $P_\rho (k)\sim k^1,$ but
compared to adiabatic, scale-free perturbations
the turnover to $P_\rho (k)\sim k^{-3}$ power
law behavior on small scales occurs on a larger scale
for the isocurvature case.
For the amplitude of density perturbations 
normalized on small scales (for example, using
$\sigma _8),$ this gives about thirty times more
power in the matter perturbations 
on large scales and entails an excessive
CMB anisotropy on large angles. This work
has been extended to consider mixtures of 
CDM isocurvature and adiabatic fluctuations,
low density universes, and more recent CMB
data \cite{stompor,gouda}.

The neutrino isocurvature modes discussed here allow 
for a spatially varying relative density of photons and neutrinos and
a relative velocity between the photon and 
neutrino components as well. For the neutrino isocurvature density 
mode, the total density perturbation vanishes but the
relative density of neutrinos and photons varies
spatially. On superhorizon scales, the neutrinos and photons
evolve similarly but upon entering the horizon
the neutrinos free stream, developing nonvanishing
higher moments of the neutrino phase 
space density $F_{\nu \ell } (k),$ while 
because of Thomson scattering the photons continue
to behave much as a perfect fluid.
These distinct behaviors subsequently 
generate perturbations in the total density.
For the neutrino isocurvature velocity mode, the 
rest frames of the neutrino and the photon fluids do
not coincide. The relative velocities are such 
that initially the perturbation in the total
momentum density  vanishes. If this last condition
were not satisfied, the metric perturbation
generated by this mode would diverge at early 
times, rendering the mode a singular mode, which
according to the discussion above 
should be excluded. But the lack of divergence
owing to this cancellation makes
this an admissible regular perturbation mode. 
The neutrino isocurvature mode solutions are implicit in 
the work of Rebhan and Schwarz\cite{nnewb} and of
Challinor and Lasenby\cite{nnewc}, but their implications 
were not explored.

A possible obstacle to exciting neutrino isocurvature modes
arises at early times, from processes in which neutrinos are
generated or scattered. If the neutrino chemical potentials 
vanish, so that there are in each generation 
precisely as many neutrinos as antineutrinos,
a spatially varying relative density can only be
established at a temperature sufficiently low
that the processes turning photons into 
neutrino-antineutrino pairs and vice versa
had already been frozen out---that is, below a few MeV.
Nonvanishing chemical potentials for the various 
neutrino species can protect variations in the 
ratio $(\rho _\nu /\rho _\gamma )$ from 
erasure by processes involving $\nu \bar \nu $
annihilation, and observational
constraints that would rule out neutrino chemical
potentials of the required magnitude are lacking. 
$L$-violating processes mediated through sphalerons, 
unsuppressed at temperatures above the 
electroweak phase transition, can readjust the 
neutrino chemical potentials
$\mu _{\nu _e},$ $\mu _{\nu _\mu },$
and $\mu _{\nu _\tau }$ and, moreover, 
can convert lepton asymmetries in the neutrino sector into 
baryon asymmetry, but if the neutrino chemical potentials
satisfy $\mu _{\nu _e}+\mu _{\nu _\mu }+\mu _{\nu _\tau }=0,$
then no tendency favoring
sphalerons over anti-sphalerons is introduced and the net effect 
of these electroweak processes vanishes. Recall that the 
neutrino overdensity is proportional to 
${\mu _{\nu _e}}^2+{\mu _{\nu _\mu }}^2+{\mu _{\nu _\tau }}^2.$
Similarly, for the neutrino isocurvature velocity mode 
scattering of neutrinos with other
components dampens this mode at very early times, and
for this mode to be relevant there must exist a process capable of
exciting it after this dampening effect has frozen out.

There exist many candidate mechanisms that might generate these 
neutrino isocurvature modes. The neutrino density isocurvature
mode could be generated during inflation if the theory included a 
light scalar field carrying lepton number, with mass much smaller 
than the Hubble constant during inflation.\cite{nnewd}
During inflation this field would be excited but would contribute
negligibly to the density of the universe. After inflation, 
when the Hubble constant fell below the mass of the scalar field
it would oscillate and decay, producing a lepton asymmetry. We
would expect the neutrino chemical potential to be 
proportional to the scalar field value, so the most natural possibility
would be Gaussian, scale invariant perturbations in the neutrino chemical
potential, with the density perturbation in the neutrinos being proportional
to the square of the chemical potential.

The neutrino velocity mode could have been excited by the decay of
relics such as cosmic strings, walls, or superstrings, surviving
until after the neutrinos had decoupled and then decaying into
neutrinos. The neutrino fluid produced in such processes would 
be perturbed both in its density and 
velocity, and these perturbations would be isocurvature in 
character \cite{ltter}.
Another possibility for exciting the modes arises from magnetic
fields frozen into the plasma whose stress gradients impart a 
velocity to the photon-lepton-baryon plasma but not to the 
neutrino component.

\def\x{{\bf x}}
\def\k{{\bf k}}
\def\H{{\cal H}}

\section{Identifying the Modes}

We now turn to identifying the possible perturbation 
modes using synchronous gauge with the 
line element 
\ba 
ds^2=a^2(\tau )\cdot \Bigl[ 
-d\tau ^2+\bigl( \delta _{ij}+h_{ij}\bigr) ~dx^idx^j~
\Bigr] .
\ea
For spatial dependence of a given wavenumber $\k ,$ we define 
\ba
h_{ij}(\k ,\tau)=e^{i\k \cdot \x }\cdot 
\left[ \hat k_i~ \hat k_j ~h(\k ,\tau)+
\left(  \hat k_i~ \hat k_j-{1\over 3} \delta _{ij}\right) 
6\eta (\k ,\tau) \right] . 
\ea
Our discussion generally follows the 
notation of ref. \cite{ma}. 

The linearized Einstein equations are
\ba
k^2\eta -{1\over 2}\H \dot h &=&
{~}(4\pi Ga^2)~\delta {T^0}_0,\nonumber\\ 
k^2\dot \eta &=&
{~}(4\pi Ga^2)~(\bar \rho +\bar p)\theta ,\nonumber\\
\ddot h+2\H \dot h-2k^2\eta &=& -(8\pi Ga^2)~\delta {T^i}_i,\nonumber\\ 
\ddot h+6\ddot \eta +2\H (\dot h+6\dot \eta )-2k^2\eta &=&
-(24\pi Ga^2)(\bar \rho +\bar p)\sigma 
\ea
where we define  $(\bar \rho +\bar p)\theta =ik^j~{T^0}_j$
and $(\bar \rho +\bar p)\sigma =-(\hat k_i\hat k_j-{1\over 3}
\delta _{ij}){T_i}^j.$
We define $\H =\dot a/a.$
With only a single fluid, $\theta $ is simply the
divergence $(\nabla \cdot {\bf v}).$

Assuming $N$ fluids, labeled $(J=1,\ldots ,N),$ one may
rewrite the above as 
\ba
k^2\eta -{1\over 2}\H \dot h &=&
-{3\over 2}\H ^2\sum _J\Omega _J\delta _J,\nonumber\\
k^2\dot \eta &=&
{~~}{3\over 2}\H ^2\sum _J\Omega _J(1+w_J)\theta _J,\nonumber\\
\ddot h+2\H \dot h-2k^2\eta &=& -9\H ^2\sum _J\Omega _J~c_{sJ}^2~\delta _J,\nonumber\\
\ddot h+6\ddot \eta +2\H (\dot h+6\dot \eta )-2k^2\eta &=&
-12\H ^2\Omega _\nu \sigma _\nu 
\label{apad}
\ea
where $w_J={p_J/\rho _J}$ and $c_{s}^2=\partial p_J/\partial \rho _J.$
In the last equation only the neutrino contribution to the 
anisotropic stress is included. At early times only the neutrino
contribution is relevant, but later when the photons and baryons begin 
to decouple the photon quadrupole moment must also be included.

For the photons, the equations of motion at early times 
before the baryonic component of the fluid was significant are
\ba
&&\dot \delta _\gamma +{4\over 3}\theta _\gamma +{2\over 3}\dot h=0,\nonumber\\
&&\dot \theta _\gamma -{1\over 4}k^2\delta _\gamma =0.
\ea
Similarly, for the neutrinos, after neutrino decoupling at temperatures
of $\sim$ 1 MeV, we have  
\ba
&&\dot \delta _\nu +{4\over 3}\theta _\nu +{2\over 3}\dot h=0,\nonumber\\
&&\dot \theta _\nu -{1\over 4}k^2\delta _\nu +k^2\sigma _\nu =0,\nonumber\\ 
&&\dot \sigma _\nu ={2\over 15}
\left[ 2\theta _\nu +\dot h+6\dot \eta \right] -{3\over 10} k F_{\nu 3}.
\ea
where $\sigma_\nu = F_{\nu 2}/2$ is the quadrupole moment of the neutrino
phase space density and $F_{\nu l}$ is the $l$-th multipole, 
in the notation used below. 

The photon and neutrino evolution equations above 
differ only in the
presence of an anisotropic stress term $\sigma_\nu.$ Photons, 
because of their frequent scattering by charged leptons 
and baryons, at early times are unable to develop 
a quadrupole moment in their velocity distribution. 
Neutrinos, on the other hand, develop significant
anisotropic stresses upon crossing the horizon. 
The addition of an extra degree of freedom that would result 
from the equation for 
$\dot \sigma _\nu $ is avoided by setting 
$\sigma _\nu=0$ at $\tau =0.$\footnote{If we were to 
consider
the physical decaying mode with $\eta \propto \tau ^{-1},$ we
would find 
this condition cannot be satisfied. As discussed above, 
decaying modes are inevitably associated with the 
breakdown of the 
fluid approximation at early times.}

For a CDM component, the equations of motion are
\ba
&&\dot \delta _c+\theta _c+{1\over 2}\dot h=0,\nonumber\\
&&\dot \theta _c+\H \theta _c=0.
\ea

At later times, when $\Omega _b$ becomes comparable to
$\Omega _\gamma ,$ the photon equations of motion 
are modified as follows:
\ba
\dot \delta _\gamma 
+{4\over 3}\theta _\gamma +{2\over 3}\dot h&=&0,\nonumber\\
\dot \theta _\gamma -{k^2\over 4}\delta _\gamma +an_e\sigma _T
(\theta _\gamma -\theta _b)&=&0.
\label{apagaa}
\ea
Here $a$ is the scale factor, $n_e$ is the electron density, and
$\sigma _T$ is the Thomson scattering cross section.
The baryon equations of motion are
\ba
\dot \delta _b
+\theta _b+{1\over 2}\dot h&=&0,\nonumber\\
\dot \theta _b+\H \theta _b +{4\over 3}{\Omega _\gamma \over \Omega _b}
an_e\sigma _T
(\theta _b-\theta _\gamma )&=0&.
\label{apagbb}
\ea
At early times, the characteristic time for the synchronization of the 
photon and baryon velocities $t_{b\gamma }\approx 1/(n_e\sigma _T)$
is small compared to the expansion time $t_{exp}\approx a\tau $ and to the 
oscillation period $t_{osc}\approx (a\tau /k).$ (We use units where
the speed of light $c=1$.)
Any deviation of
$(\theta _\gamma -\theta _b)$ from zero rapidly decays away.
This may be seen by subtracting the second equation of \ref{apagbb}~from 
that of \ref{apagaa}~and regarding 
$\H \theta _b+{1\over 3}k^2\delta _\gamma $ as a forcing 
term. In the limit $\sigma _T\to \infty ,$ one obtains 
$\theta _b=\theta _\gamma $---in other words, a tight coupling 
between the baryons and the photons. Therefore, 
at early times we may set $\theta _b=\theta _\gamma =
\theta _{\gamma b}.$ 

In the tight coupling approximation the evolution 
equation for $\theta _{\gamma b}$ is obtained by 
adding the second equation of \ref{apagaa}~multiplied by
${4\over 3}\Omega _\gamma $ to the second equation of \ref{apagbb} ~multiplied by
$\Omega _b,$ so that the scattering terms cancel giving
\ba
\left( {4\over 3}\Omega _\gamma +\Omega _b\right)
\dot \theta _{\gamma b}
=-\Omega _b\H\theta _{\gamma b}+{1\over 3}\Omega _\gamma k^2\delta _\gamma 
\ea
and the baryon and photon density contrasts evolve according to
\ba
\dot \delta _b&=&-\theta _{\gamma b}-{1\over 2}\dot h,\nonumber\\
\dot \delta _\gamma &=&-{4\over 3}\theta _{\gamma b}-{2\over 3}\dot h.
\ea
In the absence of baryon isocurvature perturbations, 
$\delta _\gamma ={4\over 3}\delta _b,$
making one of these equations redundant, but both equations are required
for the most general type of perturbation. 

Although an excellent approximation early on, the 
tight coupling assumption breaks down at later times 
as the photons and baryons decouple, and for a more accurate
description the two equations in \ref{apagaa}~for the 
time derivatives of the monopole and dipole moments 
of the velocity of the photon phase space distribution 
must be replaced by the following 
infinite hierarchy of equations for the time
derivatives of higher order moments
of the photon distribution function as well:
\ba
\dot \delta _\gamma &=&
-{4\over 3}\theta _\gamma -{2\over 3}\dot h,\nonumber\\ 
\dot \theta _\gamma &=&k^2\left( {\delta _\gamma \over 4}
-{F_{\gamma 2}\over 2}\right)
+an_e\sigma _T(\theta _b-\theta _\gamma ),\nonumber\\
\dot F_{\gamma 2}&=&
{8\over 15}\theta _\gamma -{3\over 5}kF_{\gamma 3}+
{4\over 15}\dot h
+{8\over 5}\dot \eta 
-{9\over 10}an_e\sigma _TF_{\gamma 2},\nonumber\\
\dot F_{\gamma \ell }&=&
{k\over 2\ell +1}\left[ {\ell }F_{\gamma (\ell -1)} -
{(\ell +1)}F_{\gamma (\ell +1)}
\right] -an_e\sigma _TF_{\gamma \ell }
\ea
where $\ell \ge 3$ for the last equation. 
Initially, as $\tau \to 0,$ $F_{\gamma \ell }=0$ 
for $\ell \ge 2.$ If this condition were relaxed
an infinite number of decaying modes would appear. 
The fluid approximation for the photons, eqn.~\ref{apagaa},
is obtained using only the first two of the above
equations combined with the approximation $F_{\gamma 2}=0.$

To describe the neutrinos during and after horizon
crossing requires a Boltzman hierarchy for
$\delta _\nu ,$ $\theta _\nu ,$ $F_{\nu \ell }$
identical to the one above except that the Thomson 
scattering terms are omitted. This assumes that
neutrino masses are irrelevant. 

Before solving the Einstein equations
\ref{apad}, we first identify the two gauge modes
resulting from the residual gauge freedom remaining
within synchronous gauge. A general infinitesimal
gauge transformation considered to linear order 
is a coordinate transformation $x^\mu \to 
x^{\prime \mu }(x)$ where $x^\mu =x^{\prime \mu }+
\epsilon ^\mu (x).$ From the transformation 
rule for tensors 
\ba
{g^\prime }_{\mu \nu }(x')=
{\partial x^\xi \over \partial x^{\prime \mu }}
{\partial x^\eta \over \partial x^{\prime \nu }}
g_{\xi \eta }(x),
\ea
it follows that
\ba
\delta g_{\mu \nu }=\epsilon ^\xi ~{\partial g_{\mu \nu }^{(0)}\over 
\partial x^\xi }+{\epsilon ^\xi }_{,\mu }~g_{\xi \nu }^{(0)}
+g_{\mu \xi }^{(0)}~{\epsilon ^\xi }_{,\nu }
\label{apai}
\ea 
where $g_{\mu \nu }^{(0)}$ is the unperturbed, zeroth-order
metric. Applying \ref{apai}~to the metric
$ds^2=a^2(\tau ) \Bigl[ -(1-h_{00})d\tau ^2+2h_{0i}~d\tau dx^i
+(\delta _{ij}+h_{ij})dx^idx^j\Bigr] ,$ where 
\ba
\tau =\tau ^\prime +T(\x ,\tau ), \quad 
\x =\x '+{\bf S}(\x ,\tau )
\ea
where $S$ and $T$ are regarded as infinitesimal, one obtains
\ba
\delta h_{00}&=&-{2\dot a\over a}T(\x ,\tau )-2\dot T(\x ,\tau ),\nonumber\\
\delta h_{0i}&=&-T_{,i}(\x ,\tau )+\dot S_i(\x ,\tau ),\nonumber\\
\delta h_{ij}&=&S_{i,j}+S_{j,i}+{2\dot a\over a}\delta _{ij}T. 
\ea
For the density contrast for a perfect fluid component 
labeled by $\alpha ,$ one obtains $\delta (\delta _\alpha )
=-3(1+w_\alpha )\H T.$ After a gauge transformation
the velocity is shifted by $\delta {\bf v}=-\dot S(\x ,\tau )$
for all components.

The synchronous gauge condition $\delta h_{00}=0$ implies 
that $T(\x ,\tau )=A(\x )/a(\tau ),$ and $\delta h_{0i}=0$
implies that
\ba
{\bf S}(\x ,\tau )=B(\x )+\nabla A(\x )\int {d\tau \over a(\tau )}.
\ea
Therefore
\ba
\delta h_{ij}={2\dot a\over a^2}A(\x )\delta _{ij}+2A_{,ij}
\int {d\tau \over a(\tau )}+B_{i,j}+B_{j,i}.
\ea

For a radiation-dominated universe [with $a(\tau )=\tau $] 
considering for the moment only a single wavenumber, we find that%
\footnote{Note that the gauge mode $A$
disagrees with eqns.~[94] and [95] of ref.~\cite{ma},
which are incorrect.}
\ba
\delta h_{ij}=\left[ \left( {2\over \tau ^2}\delta _{ij}
-2k_ik_j\ln (\tau )\right) A(\k )+
2B~k_ik_j \right] e^{i\k \cdot \x },
\ea
or 
\ba
h&=&A\left[ {6\over \tau ^2}-2k^2\ln (\tau )\right] +2B,\quad 
\eta =A\left[ {-1\over \tau ^2}\right].
\label{apao}
\ea

To obtain a more accurate power series expansion close to
the time of matter-radiation equality and to consider the baryon and
CDM isocurvature modes, we assume a scale factor evolution
for a universe filled with matter and radiation: 
$a(\tau )=\tau +\tau ^2.$ At matter-radiation equality
$a_{eq}=1/4$ and $\tau _{eq}=(\sqrt{2}-1)/2.$ Since 
$\H =(2\tau +1)/(\tau ^2+\tau )$ has a pole at
$\tau =-1,$ the resulting power series solutions are expected
to diverge beyond the unit circle. 

For the matter-radiation universe, with $a(\tau )=\tau +\tau ^2$
eqn. \ref{apao}~is modified to become 
\ba
h&=&A\left[ {6(1+2\tau )\over \tau ^2(1+\tau )^2}
-2k^2\ln \left({\tau \over 1+\tau }\right) \right] +2B,\quad
\eta =-A\left[ {(1+2\tau )\over \tau ^2(1+\tau )^2}\right].
\ea

We now present the power series solutions. 
$R_\gamma $ and $R_\nu $ represent the fractional contribution
of photons and neutrinos to the total density at early times
deep within the 
radiation dominated epoch. (We ignore possible effects due to
nonvanishing neutrino masses). We also assume that we are 
considering the perturbations after the 
annihilation of electrons and positrons, so that the latter have 
dumped their energy into the photon background. 
For $N_\nu$  species of massless neutrinos we define 
$R ={7\over 8} N_{\nu} ({4\over 11})^{4\over 3}$ and we have
$R_\gamma = (1+R)^{-1}$ and $R_\nu =R(1+R)^{-1}.$ 

\vskip 25pt 
\noindent
{\it Adiabatic Growing Mode.}

\ba
h&=&{1\over 2}k^2\tau ^2,\nonumber\\
\eta &=& 1-{5+4R_\nu \over 12(15+4R_\nu )}k^2\tau ^2,\nonumber\\
\delta _c&=&-{1\over 4}k^2\tau ^2,\nonumber\\
\delta _b&=&-{1\over 4}k^2\tau ^2,\nonumber\\
\delta _\gamma &=&-{1\over 3}k^2\tau ^2,\nonumber\\ 
\delta _\nu &=&-{1\over 3}k^2\tau ^2,\nonumber\\ 
\theta _c&=&0,\nonumber\\
\theta _{\gamma b}&=&-{1\over 36}k^4\tau ^3,\nonumber\\
\theta _\nu &=&-{1\over 36}\left[ {23+4R_\nu \over 15 +4R_\nu }\right] 
k^4\tau ^3,\nonumber\\
\sigma _\nu &=&{2\over 3(12+R_\nu )}k^2\tau ^2.
\ea

\vskip 25pt
\noindent
{\it Baryon Isocurvature Mode.}

\ba
h&=&4\ob \tau -6\ob \tau ^2,\nonumber\\
\eta &=&-{2\over 3}\ob \tau +\ob \tau ^2 ,\nonumber\\
\delta _c&=&-2\ob \tau +3\ob \tau ^2,\nonumber\\
\delta _b&=&1-2\ob \tau +3\ob \tau ^2,\nonumber\\
\delta _\gamma &=&-{8\over 3}\ob \tau +4\ob \tau ^2,\nonumber\\
\delta _\nu &=&-{8\over 3}\ob \tau +4\ob \tau ^2,\nonumber\\
\theta _c&=&0,\nonumber\\
\theta _{\gamma b}&=&-{1\over 3}\ob k^2\tau ^2,\nonumber\\
\theta _\nu &=&-{1\over 3}\ob k^2\tau ^2,\nonumber\\
\sigma _\nu &=&{-2\ob \over 3(2R_\nu +15)}k^2 \tau ^3.
\ea
There is no regular baryon velocity mode because of the 
tight coupling of the baryons to the photons.

\vskip 25pt 
\noindent
{\it CDM Isocurvature Mode.}

\ba
h&=&4\oc \tau -6\oc \tau ^2,\nonumber\\
\eta &=&-{2\over 3}\oc \tau +\oc \tau ^2 ,\nonumber\\
\delta _c&=&1-2\oc \tau +3\oc \tau ^2,\nonumber\\
\delta _b&=&-2\oc \tau +3\oc \tau ^2,\nonumber\\
\delta _\gamma &=&-{8\over 3}\oc \tau +4\oc \tau ^2,\nonumber\\
\delta _\nu &=&-{8\over 3}\oc \tau +4\oc \tau ^2,\nonumber\\
\theta _c&=&0,\nonumber\\
\theta _{\gamma b}&=&-{1\over 3}\oc k^2\tau ^2,\nonumber\\
\theta _\nu &=&-{1\over 3}\oc k^2\tau ^2,\nonumber\\
\sigma _\nu &=&{-2\oc \over 3(2R_\nu +15)}k^2 \tau ^3.
\ea
The CDM velocity mode may be identified with the gauge mode. 
This can be seen from the equation for the CDM
velocity which is decoupled from the other equations.
$\theta _{CDM}$ behaves as $\tau ^{-1}$ at early times.
If there were several species of CDM, however, a new 
physical, non-gauge relative velocity mode would arise,
which would be divergent at early times. 

\vskip 25pt
\noindent

{\it Neutrino Isocurvature Density Mode.}

\ba
h&=&{\ob R_\nu \over 10R_\gamma }k^2\tau ^3,\nonumber\\
\eta &=&-{R_\nu \over 6(15+4R_\nu )}k^2\tau ^2,\nonumber\\
\delta _c&=&{-\ob R_\nu \over 20R_\gamma }k^2\tau ^3,\nonumber\\
\delta _b&=&{1\over 8}{R_\nu \over R_\gamma }k^2\tau ^2,\nonumber\\
\delta _\gamma &=&-{R_\nu \over R_\gamma }
+{1\over 6}{R_\nu \over R_\gamma }k^2\tau^2,\nonumber\\
\delta _\nu &=&1-{1\over 6}k^2\tau ^2,\nonumber\\
\theta _c&=&0,\nonumber\\
\theta _{\gamma b}&=&-{1\over 4}{R_\nu \over R_\gamma }k^2\tau +
{3\ob R_\nu \over 4R_\gamma ^2}k^2\tau ^2,\nonumber\\
\theta _\nu &=&{1\over 4}k^2\tau ,\nonumber\\
\sigma _\nu &=&{1\over 2(15+4R_\nu )}k^2\tau ^2.
\ea
Physically, one starts with a uniform energy density,
with the sum of the neutrino and photon densities unperturbed.
When a mode enters the horizon, the photons behave
as a perfect fluid, while the neutrinos free stream,
creating nonuniformity in the energy density, pressure, 
and momentum density, thus sourcing
metric perturbations.

\vskip 25pt
\noindent
{\it Neutrino Isocurvature Velocity Mode.}

\ba
h&=&{3\over 2}\ob {R_\nu \over R_\gamma }k\tau ^2,\nonumber\\
\eta &=&-{4R_\nu \over 3(5+4R_\nu )}k\tau +
\left( {-\ob R_\nu \over 4R_\gamma }+{20R_\nu \over 
(5+4R_\nu )(15+4R_\nu )}\right) k\tau ^2,\nonumber\\
\delta _c&=&-{3\ob \over 4}{R_\nu \over R_\gamma }k\tau ^2,\nonumber\\
\delta _b&=&{R_\nu \over R_\gamma }k\tau -
{3\ob R_\nu (R_\gamma +2)\over 4R_\gamma ^2}k\tau ^2,\nonumber\\
\delta _\gamma &=&{4\over 3}{R_\nu \over R_\gamma }k\tau -
{\ob R_\nu (R_\gamma +2)\over R_\gamma ^2}k\tau ^2,\nonumber\\
\delta _\nu &=&-{4\over 3}k\tau - {\ob R_\nu \over R_\gamma }k\tau ^2,\nonumber\\
\theta _c&=&0,\nonumber\\
\theta _{\gamma b}&=&-{R_\nu \over R_\gamma }k+{3\ob R_\nu \over R_\gamma ^2}k\tau +
{R_\nu \over R_\gamma }\left(
{3\ob \over R_\gamma }-{9\ob ^2\over R_\gamma ^2}
\right) k\tau ^2+
{R_\nu \over 6R_\gamma }k^3\tau ^2,\nonumber\\
\theta _\nu &=&k-{(9+4R_\nu )\over 6(5+4R_\nu )}k^3\tau ^2,\nonumber\\
\sigma _\nu &=&{4\over 3(5+4R_\nu )}k\tau 
+{16R_\nu \over (5+4R_\nu )(15+4R_\nu )}k\tau ^2,\nonumber\\
F_{\nu 3}&=&{4\over 7(5+4R_\nu )}k^2\tau ^2.
\ea
Here the neutrinos and photons start with uniform total
density and uniform density ratio but with relative velocities
matched so that initially the total momentum density vanishes.
If the relative momenta were not perfectly matched, the metric
perturbation generated would diverge at early times, as in the 
adiabatic decaying mode. But because of the perfect match, a
divergence at early times is avoided. 

It is also possible, at least in principle, to consider
regular modes with higher moments of 
$F_{\nu \ell }$ with $\ell \ge 3$ excited initially,
as was considered in ref. \cite{nnewb}; however, it is difficult
to envision plausible mechanisms for exciting these higher
moment modes. 

\vskip 14pt
\noindent
{\bf Newtonian Potentials.} In this paper we have used
synchronous gauge, but for completeness we give the 
form of the Newtonian potentials for the regular modes
presented above. The Newtonian potentials $\phi $ and 
$\psi $ are related to the synchronous variables as
follows:
\ba
\phi &=&{1\over 2k^2}\bigl[ \ddot h+6\ddot \eta  
+\H(\dot h+6\dot \eta )\bigr] ,\nonumber\\
\psi &=& \eta -{\H \over 2k^2}[\dot h+6\dot \eta ].
\ea
We define the Newtonian potentials according to the 
convention $ds^2=a^2(\tau )[-d\tau ^2(1+2\phi )+
dx^i~dx^j~\delta _{ij}(1-2\psi )].$
We now give the Newtonian potentials to leading order.
For the growing adiabatic mode 
\ba
\phi &=&{10\over (15+4R_\nu )},\nonumber\\
\psi &=&{10\over (15+4R_\nu )}.
\ea
For the neutrino isocurvature density mode 
\ba
\phi &=&{-2R_\nu \over (15+4R_\nu )},\nonumber\\
\psi &=&{R_\nu \over (15+4R_\nu )}. 
\ea
For the neutrino isocurvature velocity mode 
\ba
\phi &=&{-4R_\nu \over (15+4R_\nu )}k^{-1}\tau ^{-1},\nonumber\\
\psi &=&{4R_\nu \over (15+4R_\nu )}k^{-1}\tau ^{-1}.
\ea
The potentials for the baryon 
isocurvature mode are
\ba
\phi &=&{(4R_\nu -15)\ob \over 2(15+2R_\nu )}\tau ,\nonumber\\
\psi &=&{-(4R_\nu -15)\ob \over 6(15+2R_\nu )}\tau ,
\ea
and for the CDM isocurvature mode are
\ba
\phi &=&{(4R_\nu -15)\oc \over 2(15+2R_\nu )}\tau ,\nonumber\\
\psi &=&{-(4R_\nu -15)\oc \over 6(15+2R_\nu )}\tau .
\ea
It is curious that the Newtonian potential diverges
at early times for the neutrino isocurvature velocity mode
while in synchronous gauge there is no singularity. 
It appears that synchronous gauge is a more physical
gauge and that Newtonian gauge is inadequate for modes 
based on anisotropic stresses. The dimensionless 
Ricci curvature $a^2(\tau )\tau ^2R$ is nonsingular at early times.
In any case, synchronous gauge is more physical in the sense that
its evolution is {\it local} whereas in Newtonian gauge the evolution
of the shape of the constant cosmic time hypersurfaces depends on how 
matter behaves infinitely far away (because the gauge choice is defined in
terms of the {\it scalar-vector-tensor} decomposition,
which is nonlocal). The divergence
of the Newtonian potentials arises from the need to warp the surfaces
of constant cosmic time so as to put the metric in a spatially
isotropic form; however, there is nothing at all physical about this 
particular form. The neutrino isocurvature velocity mode
was excluded in ref. \cite{nnewb} because of the behavior of the 
Newtonian potentials at early times; however, when a physical
phenomemon can be described in a nonsingular manner in some gauge, 
its singularity in another gauge should be regarded as a 
coordinate singularity. 

\begin{figure}
\centerline{\psfig{file=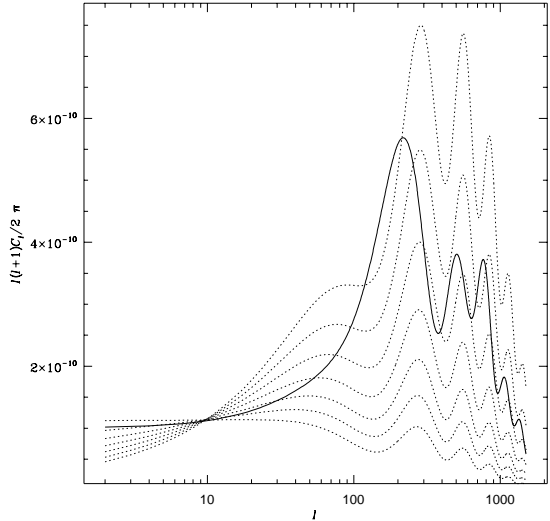,width=3.in}}
\caption{
{\bf CMB Anisotropy for the Neutrino Isocurvature Density Mode.}
We plot $\ell (\ell +1)c_\ell /2\pi $ for the 
neutrino isocurvature density mode 
(the dashed curves) for initial power spectra 
$P_{\delta _\nu }\sim k^\alpha $ where $\alpha =-3.0,\ldots ,-2.4$
increasing in increments of $0.1$ from bottom to top at
large $\ell .$ The adiabatic growing mode (the solid curve) with a 
scale-invariant spectrum is included for comparison.
All curves are normalized to COBE. For the lowest curve the 
variations in the photon-to-neutrino ratio
obey a scale-invariant initial power spectrum.
}
\end{figure}

\begin{figure}
\centerline{\psfig{file=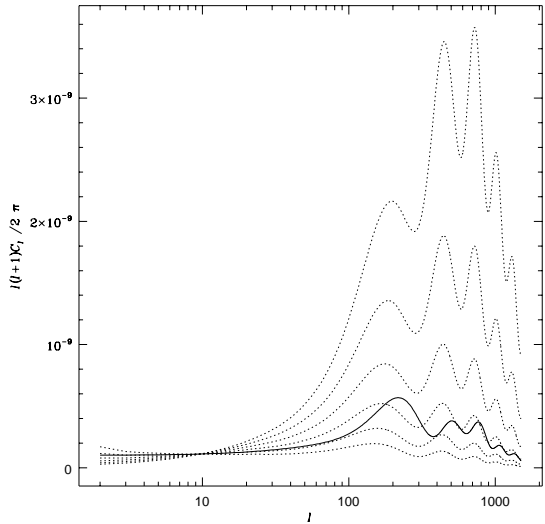,width=3.in}}
\caption{
{\bf CMB Anisotropy for the Neutrino Isocurvature Velocity Mode.}
We plot $\ell (\ell +1)c_\ell /2\pi $ for the neutrino isocurvature 
velocity mode (the dashed curves) for initial power spectra 
$P_{v_\nu }\sim k^\alpha $ with $\alpha =-3.0,\ldots ,-2.0$
increasing in increments of $0.2$ from bottom to top 
at large $\ell .$ $\alpha =-2.0$ corresponds to 
a white noise initial power spectrum in the divergence
of the velocity field, possibly resulting from a large
number of small explosions.
}
\end{figure}

\section{Discussion}

We have identified five nondecaying modes
corresponding to each wavenumber: an adiabatic growing mode, 
a baryon isocurvature mode, a CDM isocurvature mode,
a neutrino density isocurvature mode, and 
a neutrino velocity isocurvature mode. 
Under the assumption that the primordial perturbations 
are small enough so that the linear theory suffices, 
two-point derived observables are completely determined by the 
generalization of the power spectrum given by the
correlation matrix
\ba
\langle A_m(k)A_n(-k)\rangle \ea
where the indices $(m, n=1,\ldots, 5)$ label the modes,
independently of whether or not
the primordial fluctuations were Gaussian. 
In Figs.~1 and 2 the shape of the CMB moments for
the neutrino isocurvature modes are indicated. 
In this computation the cosmological parameters 
$H_0=50{\rm km~s}^{-1},$
$\Omega _b=0.05,$ $\Omega _c=0.95$ were assumed. 
In a companion paper, we examine prospects for
constraining the amplitudes of these modes using
upcoming MAP and PLANCK data.

\vskip 12pt
\noindent
{\bf Acknowledgements:} We would like to thank David Spergel 
for useful discussions at an early stage of this project. 
The CMB spectra were computed using a modified
version of the code CMBFAST written by Uros Seljak and Matias
Zaldarriaga. We would like to thank Anthony Challinor, Anthony
Lasenby, and Dominik Schwarz for useful comments.
This work was supported in part
by the UK Particle Physics and Astronomy Research 
Council. KM was supported by the Commonwealth Scholarship 
Commission in the UK.


\begin{thebibliography}{99}

\bibitem{infpert}Hawking, S.W., 1982, ``The Development of Irregularities
in a Single Bubble Inflationary Universe," Phys. Lett., 115B, 295;
Starobinsky, A.A., 1982, ``Dynamics of Phase Transition in the New
Inflationary Scenario and the Generation of Perturbations,"
Phys. Lett., 117B, 125; Guth, A. \& Pi, S.Y., 1982, 
``Fluctuations in the New Inflationary Universe,"
Phys. Rev. Lett., 49, 1110;
Bardeen, J., Steinhardt, P., \& Turner, M. 1983,
``Spontaneous Generation of Almost Scale-Free Density 
Perturbations in an Inflationary Universe,"
Phys. Rev. D., 28, 679

\bibitem{nongauss}See, for example:
Salopek, D.S., 1992,``Cold-Dark Matter Cosmology with
Non-Gaussian Fluctuations from Inflation," Phys. Rev., D45,
1139; 
Kofman , L., \&  Linde, A. 1987, ``Generation of Density
Perturbations in Inflationary Cosmology," Nucl. Phys., B282,
555;
Vishniac, E., Olive, K. \& Seckel, D. 1987, ``Cosmic Strings and
Inflation," Nucl. Phys. B, 289, 717);
Kofman, L., Linde A., \& Einasto, J. 1987 ``Cosmic Bubbles as Remnants
from Inflation," Nature, 326, 48;
Hodges, H. \&Primack, J. 1991, ``Strings, Textures and Inflation,"
Phys. Rev. D, 43, 3155;
Linde, A. 1992, ``Strings, Textures, Inflation, and Spectrum Bending,"
Phys. Lett 284B, 215 (1992);
Basu, R., Guth, A. \& Vilenkin, A. 1991, ``Quantum
Creation of Topological Defects During Inflation," Phys. Rev. D, 44,
340;
Kofman, L. 1986, ``What Initial Perturbations May Be Generated in Inflationary
Cosmological Models," Phys. Lett., 173B, 400;
Bucher, M. \& Zhu, Y. 1997,, ``Non-Gaussian 
Isocurvature Perturbations From Goldstone Modes
Generated During Inflation," Phys.Rev. D, 55, 7415;
and references therein.

\bibitem{peenew}
Peebles, P.J.E. 1999, ``An Isocurvature Cold Dark Matter Cosmogony. 
I. A Worked Example of Evolution through Inflation," Ap.J., 510, 523;
Peebles, P.J.E. 1999, ``An Isocurvature Cold Dark Matter Cosmogony. 
II. Observational Tests," Ap. J., 510, 531 

\bibitem{mapp}http://www.map.gsfc.nasa.gov 

\bibitem{plnck}http://www.astro.estec.esa.nl/SA-general/Projects/Planck/

\bibitem{paramest}See, for example:
Bond, J.R., Efstathiou, G. \& Tegmark, M. 1997, 
``Forecasting Cosmic Parameter Errors from 
Microwave Background Anisotropy Experiments,"
M.N.R.A.S., 291, L33; Knox, L. 1995, Phys. Rev. D, 53 4307;
Jungman, G., Kamionkowski, M., Kosowsky, A., 
\& Spergel, D. 1996, Phys. Rev. Lett., 76, 1007;
Zaldarriaga, M., Spergel, D.N. \& Seljak, U. 
1997, ``Microwave Background Constraints on Cosmological Parameter,"
Ap.J. 488, 1 and references therein.

\bibitem{pjepa}Peebles, P.J.E. 1987, ``Origin of the 
Large Scale Peculiar Velocity Field: A Minimal 
Isocurvature Model," Nature 327, 210;
Peebles, P.J.E. 1987, ``Cosmic Background Temperature
Anisotropy in a Minimal Isocurvature Model for Galaxy
Formation," Ap.J. 315, L73

\bibitem{bondb}Bond, J.R. \& Efstathiou, G. 1987,
``Microwave Anisotropy Constraints on Isocurvature Baryon Models,"
M.N.R.A.S., 22, 33

\bibitem{cen}Cen, R., Ostriker, J.P., \& Peebles, P.J.E. 1993,
``A Hydrodynamic Approach to Cosmology: The Primeval
Baryon Isocurvature Model," Ap.J., 415, 423

\bibitem{bond}Bond, J.R. \& Efstathiou, G. 1986, 
``Isocurvature cold dark matter fluctuations,"
M.N.R.A.S., 218, 103

\bibitem{ma}Ma, C.P., \& Bertschinger, E. 1995, ``Cosmological
Perturbation Theory in the Synchronous and Conformal
Newtonian Gauges," Ap.J., 455, 7

\bibitem{shellard}Shellard, E.P.S. \& Vilenkin, A. 1994,
{\it Cosmic Strings and Other Topological Defects,}
(Cambridge: Cambridge University Press).

\bibitem{pst}Pen, U.L., 
Seljak, U., \& Turok, N.G., 1997,
``Power Spectra in Global Defect Theories of Cosmic Structure
Formation'', 
astro-ph/9704165, Phys. Rev. Lett., 79, 1611;
Pen, U.L., Seljak, U., \& Turok, N.G., 1997,
``Polarization of the Microwave Background in Defect Models,'' 
Phys. Rev. Lett., 79, 1615;
Turok, N.G., Pen, U.L. \& Seljak, U.~1998, 
``The Scalar, Vector and Tensor Contributions to the 
CMB Anisotropies from Cosmic Defects,"
Phys.~Rev.~D, 58, 23506 (astro-ph/9706250)

\bibitem{durrer}Durrer, R., Kunz, M. \& Melchiorri, A. ~1998,
``Cosmic Microwave Background Anisotropies from Scaling Seeds: 
Global Defect Models,'' (astro-ph/9811174)

\bibitem{axenides}
Turner, M., Wilczek, F., \& Zee, A. 1983, ``Formation of Structure
in an Axion Dominated Universe," Phys. Lett. B, 125, 35;
Axenides, M., Brandenberger, R., \& Turner, M.~1983,
``Development of Axion Perturbations in an Axion Dominated
Universe," Phys. Lett., 126B, 178;
Linde, A.D. 1985, ``Generation of Isothermal Density Perturbations
in an Inflationary Universe," Phys. Lett. B, 158, 375

\bibitem{stompor}Stompor, R., Banday, A.J., \& Gorski, K.M.~1996,
``Flat Dark Matter-Dominated Models with Hybrid Adiabatic Plus
Isocurvature Initial Conditions," Ap.J., 463, 8

\bibitem{gouda}Gouda, N. \& Sugiyama, N.~1992, ``Surviving Cosmological
Models After the Discovery of Large-Angle Anisotropies in the
Microwave Background," Ap.J., 395L, 59

\bibitem{yo}Gorski, K.~\& Silk, J. 1989, ``Large Scale Microwave Background
Anisotropies in Isocurvature Baryon Open Universe Models," Ap.J., 346, 1

\bibitem{gor}Yokoyama, J. \& Suto, Y. ``Baryon Isocurvature Scenario in Inflationary
Cosmology: A Particle Physics Model and its Astrophysical Implications,"
Ap.J., 379, 427

\bibitem{sugi}Suginohara, T., \& Suto, Y.~1992, ``Large Scale Structure in Isocurvature
Baryon Models," Ap.J., 387, 431

\bibitem{chiba}Chiba, T., Sugiyama, N., \& Suto Y.~1994, ``Microwave Background Anisotropies
in Primeval Baryon Isocurvature Models: Constraints on the Cosmological Parameters,"
Ap.J., 429, 427

\bibitem{ltter}Bucher, M., Moodley, K. \& Turok, N., 1999, in preparation.

\bibitem{hu}Hu, W., Bunn, E., \& Sugiyama, N. 1995, ``COBE 
Constraints on Baryon Isocurvature Models," Ap.J., 447, 59

\bibitem{nnewa}Sugiyama, N., Sasaki, M., \& Tomita, K. 1989, 
``Cosmic Microwave Background Anisotropies and Large-Scale 
Velocity Fields in Isocurvature Hot Dark Matter Models,"
Ap. J. 338, L45
 
\bibitem{nnewb}Rebhan, A. \& Schwarz, D. 1994, ``Kinetic versus
Thermal Field Theory Approach to Cosmological Perturbations,"
Phys. Rev., D50, 2541

\bibitem{nnewc}Challinor, A. \& Lasenby, A. 1999, ``Cosmic 
Microwave Background Anisotropy in the Cold Dark Matter Model:
A Covariant and Gauge Invariant Approach," Ap. J., 513, 1 

\bibitem{nnewd}Yokoyama, J. 1994, ``Formation of Baryon Number
Fluctuation in Supersymmetric Inflationary Cosmology," 
Astroparticle Physics 2, 291

\end{thebibliography}
\end{document}